\documentclass[letterpaper]{article} 
\usepackage{aaai2026}  
\usepackage{times}  
\usepackage{helvet}  
\usepackage{courier}  
\usepackage[hyphens]{url}  
\usepackage{graphicx} 
\usepackage{natbib}  
\usepackage{caption} 
\usepackage{algorithm}

\usepackage{amsmath}
\usepackage{multirow}
\usepackage{mathtools}
\usepackage{algpseudocode}
\usepackage{nicefrac}
\usepackage{booktabs}

\usepackage{newfloat}
\usepackage{listings}
\DeclareCaptionStyle{ruled}{labelfont=normalfont,labelsep=colon,strut=off} 
\floatstyle{ruled}
\newfloat{listing}{tb}{lst}{}
\floatname{listing}{Listing}
\title{Efficient Chromosome Parallelization for \\ Precision Medicine Genomic Workflows}

\author{
    Daniel Mas Montserrat\textsuperscript{\rm 1,\rm 2}\equalcontrib,
    Ray Verma\textsuperscript{\rm 3}\equalcontrib,
    M\'{\i}riam Barrab\'es\textsuperscript{\rm 1},
    \\
    Francisco M. de la Vega\textsuperscript{\rm 1,\rm 2},
    Carlos D.\ Bustamante\textsuperscript{\rm 1,\rm 2},
    Alexander G. Ioannidis\textsuperscript{\rm 1,\rm 2,\rm 4}
}
\affiliations{
    \textsuperscript{\rm 1} Galatea Bio, 
    \textsuperscript{\rm 2} Stanford University, 
    \textsuperscript{\rm 3} New York University, 
    \textsuperscript{\rm 4} University of California, Santa Cruz
}

\usepackage{bibentry}

\begin{document}

\maketitle

\begin{abstract}
Large-scale genomic workflows used in precision medicine can process datasets spanning tens to hundreds of gigabytes per sample, leading to high memory spikes, intensive disk I/O, and task failures due to out-of-memory errors. Simple static resource allocation methods struggle to handle the variability in per-chromosome RAM demands, resulting in poor resource utilization and long runtimes. In this work, we propose multiple mechanisms for adaptive, RAM-efficient parallelization of chromosome-level bioinformatics workflows. First, we develop a symbolic regression model that estimates per-chromosome memory consumption for a given task and introduces an interpolating bias to conservatively minimize over-allocation. Second, we present a dynamic scheduler that adaptively predicts RAM usage with a polynomial regression model, treating task packing as a Knapsack problem to optimally batch jobs based on predicted memory requirements. Additionally, we present a static scheduler that optimizes chromosome processing order to minimize peak memory while preserving throughput. Our proposed methods, evaluated on simulations and real-world genomic pipelines, provide new mechanisms to reduce memory overruns and balance load across threads. We thereby achieve faster end-to-end execution, showcasing the potential to optimize large-scale genomic workflows. 
\end{abstract}


\section{Introduction}

Precision medicine seeks to personalize prevention, diagnosis, and treatment to the molecular and demographic context of each individual. Enabled by rapid declines in sequencing cost and growing clinical adoption of whole–genome sequencing (WGS), recent developments increasingly rely on genomic features that improve risk stratification and therapeutic decision-making. Two key examples are \emph{polygenic risk scores} (PRS), which aggregate measures that linearly combine the small effects of many common variants to estimate disease susceptibility, and \emph{high–resolution ancestry prediction} via local ancestry inference (LAI), which identifies the ancestral origin of chromosomal segments in admixed genomes. PRS are clinically valuable for early risk assessment, screening prioritization, and preventive interventions, while LAI improves the accuracy and fairness of genetic analyses by accounting for population structure at fine genomic scales and informing downstream interpretation. Both PRS computation and LAI can require complex, multi–stage bioinformatics pipelines that operate on WGS–scale inputs, stressing I/O and memory through variant calling, phasing/imputation, reference–panel matching, and model application.


Modern bioinformatics pipelines have grown immensely in both scale and complexity over the past decade. Today’s workflows often string together half a dozen or more distinct tools—ranging from low-level C/C++ programs for read alignment to Java-based variant callers to Python or R scripts for downstream analyses. Each of these components may have unique resource profiles and configuration parameters. At the same time, the underlying genomic datasets have increased in size: whole-genome VCFs or GVCFs generated from deep-coverage sequencing can easily exceed tens or even hundreds of gigabytes per sample. Handling such voluminous inputs without careful orchestration frequently leads to excessive disk I/O, unpredictable memory consumption, and long wall-clock times.

\begin{figure}[t]
    \centering
    \includegraphics[width=1.0\columnwidth]{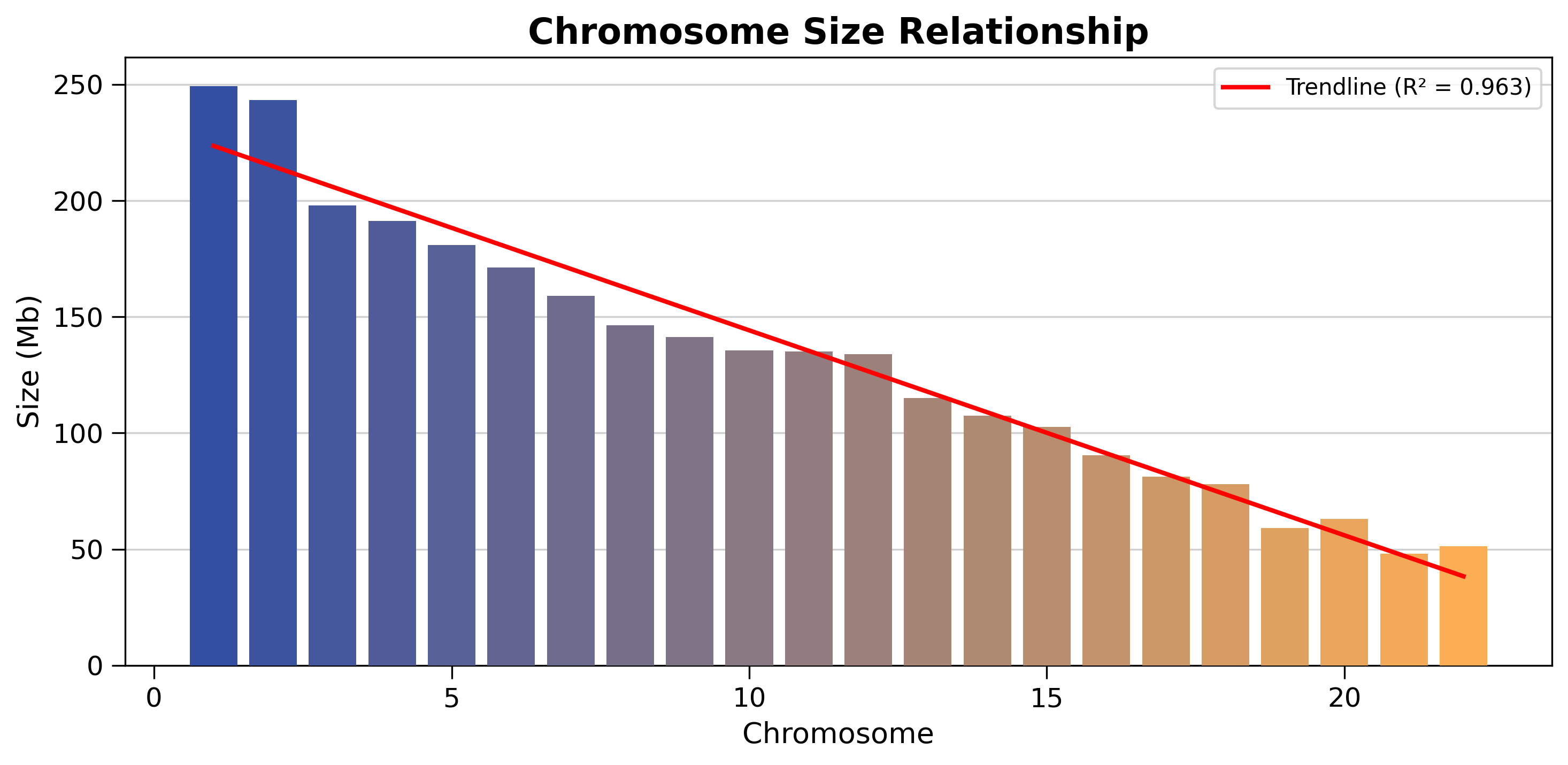}
    \caption{\textbf{Relationship between human chromosome number and its size.} 
    }
    \label{fig:chromosome_scaling}
\end{figure}


A natural strategy to manage both the data volume and memory demands is to split workloads by chromosome and parallelize them across compute resources. By treating each chromosome as an independent unit of work, one can avoid loading an entire genome’s worth of data into memory at once—mitigating out-of-memory errors when processing the largest files. Furthermore, parallel execution across chromosomes maximizes CPU utilization and reduces end-to-end latency.
Typically, the computational complexity and, consequently, the memory consumption and runtime, for processing a chromosome exhibit an almost linear relationship with the chromosome's length, itself correlated with its ordinal label (i.e., chromosome number). As shown in Figure \ref{fig:chromosome_scaling}, chromosomes vary drastically in length (for example, human chromosome 1 is roughly $4\times$ larger than chromosome 22), and so naïvely distributing one chromosome per core can lead to imbalances wherein some jobs finish quickly while others become stragglers. This scaling behavior can inform the design of resource requirement prediction and management systems.


In this work, we present three complementary systems for RAM-efficient chromosome-parallel genomics. \textbf{(1) A static scheduler} assumes a fixed concurrency budget $K$ (maximum threads) and optimizes the chromosome processing order to minimize peak memory while preserving throughput. \textbf{(2) A dynamic scheduler} treats batching as a knapsack-style packing problem: each chromosome-level job is assigned an estimated RAM cost, and the scheduler launches the jobs that maximize memory utilization, while updating estimates online based on observed peak usage with a polynomial regression. \textbf{(3) A RAM prediction module} based on symbolic regression predicts per-task memory from input characteristics (e.g., file size, software/configuration flags); these predictions plug directly into the dynamic scheduler to improve packing decisions from the outset. Together, these components balance load, curb overruns, and reduce wall-time across heterogeneous compute environments of precision medicine genomic workflows.

\section{Related Works}

\paragraph{Workflow frameworks and resource-aware scheduling.}
Many workflow orchestrators have been applied for genomic and precision medicine pipelines. General-purpose workflow engines, including GATK's Scatter--Gather, Nextflow, Snakemake, and Cromwell/WDL, among others, can support chromosome-level partitioning and per-task memory declarations \cite{ahmed2021design,van2013fastq,langer2025empowering,koster2012snakemake,voss2017full}. Specialized systems such as ADAM \& Disq for scalable formats, Toil for dynamic rescheduling, and NVIDIA Parabricks for GPU-accelerated pipelines, demonstrate adaptive resource management in practice \cite{massie2013adam,Disq2019,vivian2017toil,o2023accelerating}.

\paragraph{Machine learning for resource prediction.}
ML has long underpinned resource prediction: \cite{Matsunaga2010} introduced an SVM-based ``Predicting Query Runtime'' (PQR) for bioinformatics runtime estimation and \cite{Bankole2013} found SVMs outperform Logistic Regression (LR) and Neural Networks (NNs) for cloud CPU utilization prediction. Recent studies employ Functional Link Neural Networks for resource-usage prediction \cite{Malik2022} and compare broad model families including LR, SVM, LSTM, and Bi-LSTM for CPU forecasting, with Bi-LSTM and LR among the strongest \cite{Shaikh2024}.

\paragraph{Memory and CPU prediction for workflows.}
A parallel line of work predicts task-level runtimes and memory to avert failures and over-provisioning. HPC studies mine scheduler logs with ML to estimate job memory and enable safer submissions \cite{rodrigues2016helping}; workflow-centric systems learn task-specific, dynamic memory sizing during execution \cite{bader2024sizey} and perform online memory and time forecasting to steer provisioning \cite{bader2024ks+}. 


\paragraph{Integrating prediction into allocation and scheduling.}
Beyond standalone prediction, integration into resource managers is crucial: history-based managers for genome analysis use multivariate LR to guide allocations \cite{Badosa2019}; reinforcement learning (RL) methods dynamically select per-task CPU/RAM to reduce waste \cite{Bader2022}; and temporal-graph-network approaches enable sophisticated dynamic load balancing \cite{Rajammal2025}. At cluster scale, RL DAG schedulers such as Decima learn workload-aware policies that surpass hand-tuned heuristics, suggesting closed-loop control when coupled with telemetry and runtime feedback \cite{mao2019learning}.

\paragraph{Classical scheduling \& packing foundations.}
Casting batching/packing as a knapsack/bin-packing problem aligns with classical results: 0--1 Knapsack is NP-complete \cite{karp2009reducibility}, and First-Fit Decreasing (FFD) heuristics yield strong approximations and practical performance \cite{johnson1973near}. This foundation supports our RAM-balanced chromosome batching and justifies greedy strategies with limited look-ahead under memory constraints.


\section{Static Scheduling for Parallel Processing}
We present a simple scheduling system where chromosomes are processed in parallel with up to $K$ chromosomes simultaneously processed. In order to use resources efficiently and avoid out-of-memory errors, we optimize the order in which the chromosomes are processed based on the selected $K$.
To obtain efficient chromosome orderings, we develop a simulation-based optimization framework that models chromosome-level processing under controlled concurrency. Chromosome sizes are taken from the 1000 Genomes Project reference and used to estimate memory requirements under varying execution orders and thread counts. The model simulates the processing of each chromosome as a time-bound task, with and memory requirements proportional to chromosome size, and records instantaneous memory usage throughout the run. The number of concurrent tasks, $K$, represents the number of worker threads and ranges from $2$ to $10$ to reflect realistic high-performance computing configurations for genomics pipelines.

The optimization strategy is a black-box search procedure based on stochastic hill climbing. Starting from an initial ordering of chromosomes, the method iteratively generates new candidates by randomly swapping a small set of positions (chromosome indices). Each candidate schedule is evaluated using the simulation, and the current solution is replaced if a lower peak memory footprint is obtained. Multiple restarts ($T$) are performed to mitigate premature convergence to suboptimal local maxima. 


\subsubsection*{Optimization Formulation.}

Let $\mathcal{C}=\{1,\dots,n\}$ denote the set of chromosomes (for our application $n=22$).  
Each chromosome $i\in\mathcal{C}$ has length $\ell_i>0$ (bp). The processing time $\tau_i$ is taken to be proportional to size, $\tau_i \;=\; \eta\, \ell_i$,
with proportionality constant $\eta>0$. We assume a fixed concurrency budget of $K$ identical workers (threads).
An ordering is a permutation $\pi\in \Pi_n$, where $\pi(j)$ is the chromosome scheduled at position $j$.

Given $\pi$ and $K$, start and completion times are generated by list scheduling on identical parallel machines:


\begin{align}
s_{\pi(j)} &= 0, \qquad && j=1,\dots,K, \\
s_{\pi(j)} &= \min_{i<j} \, c_{\pi(i)}, \qquad && j=K+1,\dots,n, \\
c_{\pi(j)} &= s_{\pi(j)} + \tau_{\pi(j)}, \qquad && j=1,\dots,n.
\end{align}

Let $\mathcal{A}(t;\pi,K)=\{i\in\mathcal{C} : s_i \le t < c_i\}$ be the set of tasks active at time $t$.  
We approximate per-task memory by $m_i=\ell_i$ (any monotone size$\to$memory map can be substituted).  
Instantaneous memory and the peak over the run are:
\begin{align}
M(t;\pi,K) &= \sum_{i\in \mathcal{A}(t;\pi,K)} m_i, \\
J(\pi;K)   &= \sup_{t\ge 0} M(t;\pi,K).
\end{align}
The scheduling problem is:
\begin{equation}
\pi^\star \in \arg\min_{\pi\in S_n} \; J(\pi;K).
\end{equation}

Namely, we look for the chromosome ordering $\pi^\star$ that provides the lowest RAM peak $J(\pi;K)$, while processing $K$ chromosomes in parallel.

\paragraph{Stochastic hill-climbing search.}
Let $\pi^{(0)}$ be the initial ordering. For iteration $r=1,\dots,R$, an integer
is drawn $M_r \sim \mathrm{Unif}\{1,\dots,M_{\max}\}$ and a candidate is constructed $\tilde{\pi}^{(r)}$
by applying $M_r$ independent random swaps to $\pi^{(r-1)}$:

\begin{equation}
\begin{aligned}
\tilde{\pi}^{(r)} &= \mathsf{Swap}_{(i_1,j_1)} \circ \cdots \circ
\mathsf{Swap}_{(i_{M_r},j_{M_r})}\bigl(\pi^{(r-1)}\bigr),\\
&\qquad (i_k,j_k) \stackrel{\text{i.i.d.}}{\sim} \mathrm{Unif}\{1,\dots,n\}^2.
\end{aligned}
\label{eq:perm_update}
\end{equation}

The candidate is accepted if it improves the objective:

\begin{equation}
\pi^{(r)} \;=\;
\begin{cases}
\tilde{\pi}^{(r)}, & \text{if } J(\tilde{\pi}^{(r)};K) < J(\pi^{(r-1)};K),\\[2pt]
\pi^{(r-1)},       & \text{otherwise}.
\end{cases}
\end{equation}

To reduce sensitivity to initialization, we repeat the above for $T$ independent restarts; the reported solution for a given $K$ is:

\begin{equation}
\widehat{\pi}_K \in \arg\min_{\text{restarts } \Gamma=1,\dots,T} \; J\!\left(\pi^{(R)}_{(\Gamma)};K\right).
\end{equation}

This simple stochastic first-improvement hill-climbing scheme performs a local search over permutations. The resulting optimized orderings can be precomputed for each $K$ and used at runtime without requiring additional optimization steps, leading to a simple and easy-to-adopt parallelization scheme. The next sections describe more complex and dynamic parallelization systems.




\section{Dynamic Scheduling for Parallel Processing}

We introduce a dynamic scheduling system that maximizes RAM utilization while reducing overcommitments by continuously predicting the RAM requirements via polynomial regression and scheduling the adequate chromosomes to process at each iteration through dynamic programming. 


\subsection{RAM Prediction}

The proposed approach leverages an online predictive model that, for each chromosome $c$, estimates its resource utilization $\hat{r}_c$ for a given task. The model adaptively learns the relationship between the chromosome number $c$ and memory consumption, and is refined with new observations $r^*_c$ and previous priors on a given task. In the case where a task is over-committed, i.e., scheduled without having enough available resources leading to OOM errors and process crashes, a temporary observation is added given by $r'_c = s \cdot \hat{r}_c$, where $s \in [1, \infty]$ is a scaling parameter.

\paragraph{Polynomial Regression.}
Polynomial regression is used to learn the relationship between the chromosome number $c$ and the chromosome's required RAM. The prediction function is expressed as a polynomial with learned weights, which are determined using the least squares solution from the observed samples:


\begin{equation}
\hat{r}_c = \sum_{n=0}^{d} w_n c^{n}
\label{eq:rc}
\end{equation}

where $\hat{r}_c$ is the predicted RAM usage, $c$ is the chromosome number, $w_0, w_1, \dots, w_d$ are the learned weights, and $d$ is the preset degree of the function to be learned. Every time a new chromosome $c$ is processed, a new observation $r^*_c$ is available and a new function is fitted.

\paragraph{Conservative Bias.} We introduce an extra term $b$, which is an added bias given by the $\gamma$-th percentile of the absolute residual errors, thereby producing a conservative estimate of the resource usage. Given $O_t = \{  (r^*_i, \hat{r}_i) : 1 \leq i \leq t \}$, the set of observations and predictions at time $t$, the sorted residuals can be obtained as:

\begin{equation}
R = \{ \Delta_i = |\hat{r}_i - r^*_i| \  : \ (r^*_i, \hat{r}_i) \in O_t ,  \Delta_i \leq \Delta_{i+1} \}
\end{equation}

The conservative bias $b$ can then be computed as $b = \frac{R_{\lfloor \mu \rfloor} + R_{\lceil \mu \rceil}}{2}$,
where $ \mu = \frac{\gamma}{100} \cdot |O_t|$, $\gamma \in [0,100]
$, and $\lceil\cdot\rceil$ and $\lfloor\cdot\rfloor$ are the ceil and floor functions, respectively. 
In practice, interpolating $\gamma$ from $\gamma_{max}$ to $\gamma_{min}$ produces higher bias during early stages of scheduling when the predictor is inaccurate and lower bias later on. This decreases overcommitments at the start and increases efficiency afterwards: 

\begin{equation}
\gamma_t = \gamma_{max} - \frac{|O_t|}{|O_t|+|\bar{O_t}|}\cdot\gamma_{min}
\end{equation}

where $\gamma_t$ is the bias percentile for time $t$, $\gamma_{max}$ and $\gamma_{min}$ are parameters, and $O_t$ and $\bar{O_t}$ denote the sets of observed and unobserved tasks (chromosomes), respectively. Given the conservative bias at time $t$, the predicted RAM utilization can be computed as: $\hat{r}_{c,b,t} = \hat{r}_{c} + b_t$.

\subsection{Predictor Initialization}
Initially, when no observations $r^*_c$ are available and no prior knowledge of RAM requirements is accessible, the first $p$ tasks are run sequentially to initialize the polynomial regression predictor. Due to the lack of observed samples, the early accuracy of the regressor may be low, significantly impacting scheduling performance. 

Let $\mathcal{C}=\{1,\dots,n\}$ denote the set of chromosomes ($n=22$), $I$ be the set of tasks run sequentially, and $p$ the total number of chromosomes to process sequentially.
We investigate the following configurations of $I$ that may induce optimal scheduling early on. 

\paragraph{Biggest First.}
While this approach might initially lead to a less accurate predictor due to the lack of size diversity, it aims to achieve higher RAM utilization from the outset: 
$I = (1,2,3, \dots p)$.

\paragraph{Smallest First.}
This can lead to a quicker predictor initialization, as the smaller tasks are processed rapidly. However, initial RAM utilization might be lower compared to scheduling larger tasks: 
$I = (n, n-1, n-2, \dots (n-p))$.

\paragraph{Biggest and Smallest.}
This combined strategy aims to provide a robust initial estimate of the overall trend due to the size diversity while attaining reasonable initial memory utilization: 
$I = (1,2, \dots, \lfloor \frac{p}{2}\rfloor, n, n-1, \dots, \lceil (n- \frac{p}{2})\rceil)$.

\subsection{Task Packer}
Once $|O_t| \geq p$, the RAM usage for individual tasks is predicted using the polynomial regressor. Then, at each iteration $t$, packing algorithms are employed to maximize parallelization without exceeding available RAM. We explore two methods to generate a set $P_t\subseteq \mathcal{C}$ that contains the chromosomes to be run at time $t$ that fit in the current available RAM $a_t$:

\textbf{The Greedy method} involves maximizing the \textit{number of tasks} that fit into the available memory: 

\begin{equation}
\max_{P_t}(|P_t|) \qquad \text{subject to} \qquad \sum_{i\in P_t} r_i \leq a_t
\end{equation}

The greedy method first sorts in ascending order the pending tasks by their predicted RAM usage. Then, the algorithm iterates through the sorted list, adding each task (chromosome) to the current batch until the total predicted RAM for the batch exceeds the available memory. By prioritizing the tasks with the smallest memory footprints, this approach effectively adds the maximum number of tasks that fit into the current memory capacity.

\textbf{The Knapsack method} involves maximizing the predicted \textit{RAM utilization} at any given time: 

\begin{equation}
\max_{P_t}(\sum_{i\in P_t} r_i) \qquad \text{subject to} \qquad \sum_{i\in P_t} r_i \leq a_t
\end{equation}

The Knapsack method makes use of a sparse dynamic programming approach to maximize the total predicted RAM utilization. The algorithm iterates through each available task, building a dictionary of optimal solutions for various memory capacities up to the total available RAM. After evaluating all tasks, the algorithm selects the combination that provides the highest total predicted RAM usage under the system's available memory.

\section{RAM Prediction with Symbolic Regression}


The third proposed system takes as input a vector representation of the task to be executed, which includes the size of the input files—measured as the number of genomic variants and the number of samples—and, optionally, configuration parameters of the software being run. This information is used to predict the peak RAM usage $y$ (in MB) that the task will require during execution. In addition to the primary dataset sizes, the system can incorporate characteristics of any auxiliary data used by the software, such as reference panels, as well as algorithm-specific tuning parameters. Common machine learning models such as boosting trees or neural networks can be trained to successfully predict RAM requirements given the input features. However, such methods can increase technical debt, as infrastructure needs to be put in place to store, load, and run the trained models. Therefore, here we explore using symbolic regression, where the ML models are distilled into simple equations that can be easily deployed via one line of code while preserving good predictive performance.

\subsection{Beagle RAM Modeling}

As a concrete example, in this work we apply the method to genotype imputation using the \emph{Beagle} \cite{browning2021fast} software. Beagle is a widely used tool in genomics for phasing and imputing missing genotypes, leveraging linkage disequilibrium patterns observed in a reference panel. In this context, each task is described by a feature vector:
\[
\mathbf{x} = \left( \mathrm{Thr},\ \mathrm{Burn},\ \mathrm{Iter},\ \mathrm{Win},\ V,\ S,\ V_{\mathrm{ref}},\ S_{\mathrm{ref}} \right),
\]
where $\mathrm{Thr}$ denotes the number of CPU threads allocated to the process; $\mathrm{Burn}$, $\mathrm{Iter}$, and $\mathrm{Win}$ are Beagle-specific parameters controlling, respectively, the Markov chain Monte Carlo burn-in, the number of main iterations, and the haplotype window size; $V$ and $S$ represent the number of variants and samples in the primary dataset; and $V_{\mathrm{ref}}$ and $S_{\mathrm{ref}}$ denote the number of variants and samples in the reference panel. The target variable $y$ is defined as the peak RAM usage observed during the execution of the task.

\paragraph{Data generation.}
To construct the dataset to train and evaluate the RAM predictor model for beagle, we execute the software across a grid of configurations spanning realistic ranges for all eight features in $\mathbf{x}$. Each run produces a pair $(\mathbf{x}_i, y_i)$, where $\mathbf{x}_i$ contains the task parameters and input sizes, and $y_i$ is the measured peak RAM usage.

In the case of genotype imputation with \emph{Beagle}, the 1000 Genomes Project dataset is used as the primary source of genomic data. From this dataset, we generate a wide range of input sizes by creating subsamples containing varying numbers of variants and samples. This procedure allows us to explore the relationship between memory usage and both dimensions of dataset size. Additionally, multiple reference panels of different sizes are prepared by subsampling the same dataset, ensuring diversity in $(V_{\mathrm{ref}}, S_{\mathrm{ref}})$. The Beagle-specific parameters—\emph{Burnin}, \emph{Iterations}, \emph{Window}, and \emph{Threads}—are also varied systematically to capture the effect of software configuration on RAM requirements. 


\subsection{Symbolic Regressor}

\paragraph{Feature and label standardization.}
Before training, both the input features and the target label are standardized using statistics computed from the training set. Given the original features $\mathbf{x}_i$ and RAM usage $y_i$, the transformation is:
$
\tilde{\mathbf{x}}_i = \nicefrac{(\mathbf{x}_i - \boldsymbol{\mu}_x)}{\boldsymbol{\sigma}_x}, 
\quad 
\tilde{y}_i = \nicefrac{(y_i - \mu_y)}{\sigma_y},
$
where $\boldsymbol{\mu}_x, \boldsymbol{\sigma}_x$ denote the feature-wise means and standard deviations, and $\mu_y$ and $\sigma_y$ are the mean and standard deviation of $y$. Standardization is applied consistently across training, evaluation, and deployment. This ensures all features have comparable scales and that the learning algorithm is not biased toward high-magnitude variables. The same transformation is later inverted after prediction to recover RAM usage values in MB.

\paragraph{Teacher model: ensemble of tree-based regressors.}
As the initial “teacher” model, we train an ensemble of tree-based regressors on the standardized inputs $\tilde{\mathbf{x}}_i$ and standardized target $\tilde{y}_i$. In our implementation, the teacher is a \emph{Voting Regressor} that combines the predictions of multiple learners, specifically a Random Forest Regressor, a Histogram Gradient Boosting Regressor, and a Gradient Boosting Regressor. Each component model captures different aspects of the nonlinear dependencies and complex feature interactions present in the data, while their combination yields improved robustness and generalization performance. This ensemble achieves high predictive accuracy on the RAM usage estimation task and provides the high-fidelity reference function from which the symbolic regression model is later distilled.

\paragraph{Distillation to a symbolic regressor.}
For deployment in the workflow scheduler, we distill the ensemble teacher into a lightweight, interpretable symbolic regressor $g$. Distillation proceeds by generating a large set of synthetic input points spanning the observed feature ranges, evaluating the teacher model on these points, and fitting a symbolic expression to approximate its predictions. Symbolic regression is performed using the \texttt{PySR} \cite{cranmer2023interpretable} method, configured with a set of unary and binary operators (\texttt{abs}, \texttt{exp}, \texttt{log}, \texttt{sqrt}, etc.) and constraints on expression size and depth to control complexity. The optimization problem is:
\[
\hat{g} = \arg\min_{g \in \mathcal{G}} \sum_{i=1}^{n} \left( \hat{f}(\tilde{\mathbf{x}}_i) - g(\tilde{\mathbf{x}}_i) \right)^2 + \lambda_{\mathrm{simp}} \, \Omega(g),
\]
where $\mathcal{G}$ is the space of candidate expressions and $\Omega(g)$ penalizes overly complex formulas. 
Once trained, $g$ operates directly on standardized inputs and produces a standardized RAM prediction. The physical prediction in MB is recovered by: $\widehat{y} = g(\tilde{\mathbf{x}}) \cdot \sigma_y + \mu_y$.

\paragraph{Conformal bound for conservative scheduling.}
Once the symbolic regression model is fitted, we calibrate it to provide conservative memory usage estimates that mitigate the risk of out-of-memory (OOM) failures. This is achieved via a one-sided conformal prediction procedure. The dataset is partitioned into training and calibration subsets, with the latter used exclusively for bounding. For each calibration instance, we compute the residual between the observed peak RAM and the model’s prediction. From the empirical distribution of these residuals, we extract the $(1-\alpha)$-quantile corresponding to a desired miscoverage rate $\alpha$. The given bound is not applied as a constant offset but a $(1-\alpha)$-quantile regression is computed at each valid value $y$ of predicted RAM. Namely, we construct a piecewise-linear interpolation function mapping the predicted RAM to adjusted (conservative) values based on ordered calibration pairs $(\widehat{y}_i, y_i)$. This mapping captures the heteroscedastic nature of residuals, ensuring that the bound adapts to varying predicted RAM values. The resulting conservative predictor thus maintains monotonicity and yields robust safety margins across the range of predictions.

\subsection{Deployment in the dynamic scheduler}
During deployment, the scheduler extracts, standardizes, and evaluates the task’s features with the symbolic regression function $g$ to produce a standardized RAM prediction, which is then converted back into MB via inverse scaling. The prediction is passed through the conformal projection obtaining a conservative RAM estimate. The compact form of $g$ ensures fast evaluation without adding computational overhead at the scheduling system, and the conformal bound ensures that memory allocations are conservative enough to prevent OOM failures. The predicted RAM values from symbolic regression act as priors for the dynamic scheduler, which replace the initial sequential runs described in the "Predictor Initialization" section. Such prior values allow parallelizing the chromosome processing right from the beginning, removing the inefficient initial sequential runs that reduce resource utilization and increase runtime.


\section{Experimental Results}


\begin{figure}[htbp]
    \centering
    \includegraphics[width=1.0\columnwidth]{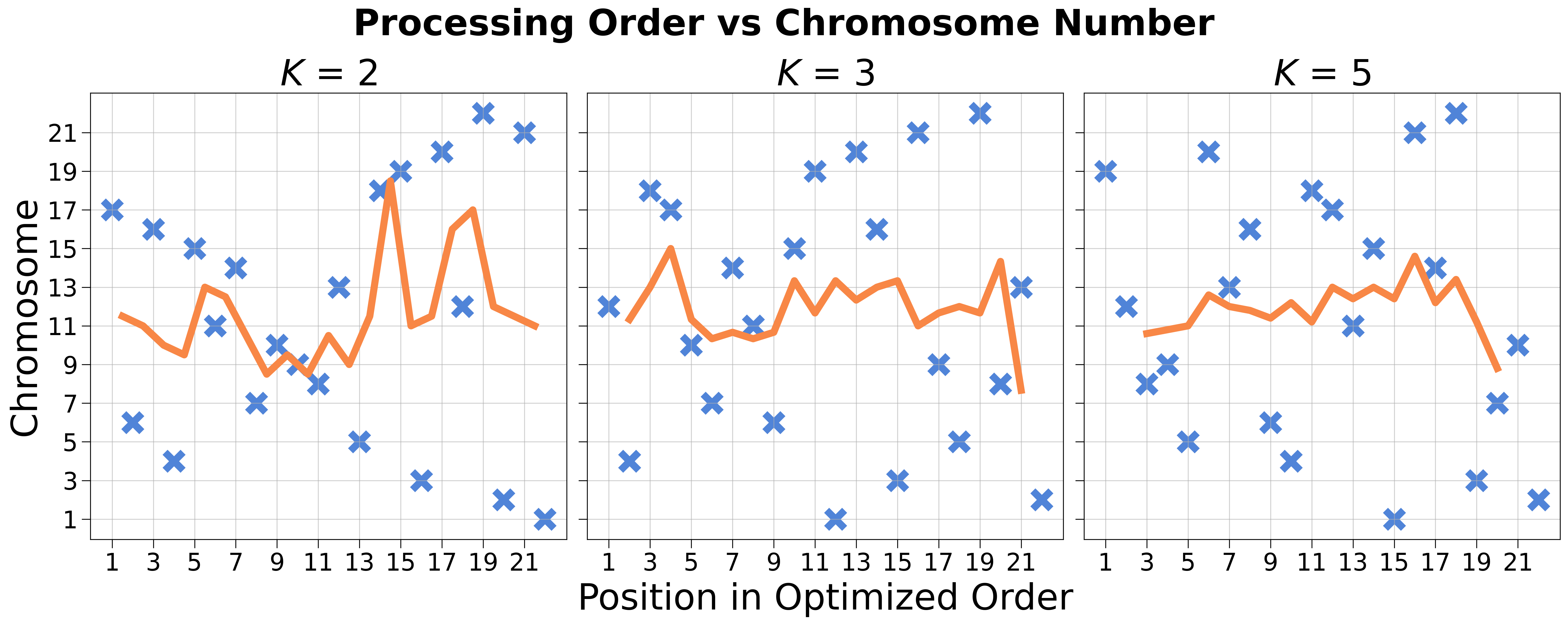} 
    \caption{\textbf{Optimized order of chromosomes} for the static scheduler for $K=2,3,5$. Each cross represents, for a given step, the chromosome being processed. A moving averaged chromosome number (orange line) indicates a balance between long and short chromosomes.}
    \label{fig:ordering}
\end{figure}

\subsection{Static Scheduler}

Analysis of the optimized orders (Fig. \ref{fig:ordering}) reveals an alternation between large and small chromosomes, an arrangement that smooths memory consumption across the execution. In the visualizations for selected $K$ values, scatter plots are overlaid with a moving average (orange line) computed within a sliding window of size $K$. Each cross represents, for each step $j$, the chromosome being processed $c_j=\pi(j)$. A moving-window average over order positions with window size $K$, $\bar{c}_u^{(K)} = \frac{1}{K} \sum_{\ell=0}^{K-1} c_{u+\ell}$, with $u=1,\dots,n-K+1$,
is plotted as the orange curve.
Empirically, $\bar{c}_u^{(K)}$ remains close to $11$ across $u$, indicating that successive windows contain a balanced mix of large (low-numbered) and small (high-numbered) chromosomes. This alternation in the optimized orders is consistent with reduced peaks in $M(t;\pi,K)$ and reduces the likelihood of high-memory peaks.

Table \ref{tab:ram_peaks} shows the peak RAM usage measured in the simulations of the sequential ordering $(1,2,\dots, 22)$ and the optimized ordering for each value of $K$. We observe a significant decrease in peak RAM usage, of up to $40\%$, particularly for low values of $K$.



\begin{table}[ht]
\centering
\caption{Peak RAM comparison: sequential vs optimized.}
\label{tab:ram_peaks}
\begin{tabular}{r r r r}
\toprule
\textbf{K} & \textbf{Sequential} & \textbf{Optimized} & \textbf{Decrease (\%)} \\
\midrule
 2 &  492.45 &  297.38 & 39.61\% \\
 3 &  690.47 &  413.47 & 40.12\% \\
 4 &  881.63 &  539.95 & 38.76\% \\
 5 & 1062.54 &  784.03 & 26.21\% \\
 6 & 1233.66 &  808.13 & 34.49\% \\
 7 & 1392.80 & 1037.98 & 25.48\% \\
 8 & 1539.16 & 1085.42 & 29.48\% \\
 9 & 1680.37 & 1186.56 & 29.39\% \\
10 & 1815.91 & 1440.64 & 20.67\% \\
\bottomrule
\end{tabular}
\end{table}

\subsection{Dynamic Scheduler}
\subsubsection{Simulation Setup.}

To overcome the time and cost limitations associated with large-scale genomic experiments, we assess the performance of our scheduler using a simulated environment. In this simulation, each chromosome-level task $i$ is assigned a "true" RAM and duration value. These values are generated from a noisy linear model to mimic real-world genomic computation. The true resource requirements for each task are calculated as follows:


\begin{equation}
\begin{aligned}
\mathrm{ram}_i &= (m i + c)\,\bigl(1 + \mathrm{uniform}(-\beta_{\mathrm{ram}}, \beta_{\mathrm{ram}})\bigr),\\
\mathrm{dur}_i &= (m i + c)\,\bigl(1 + \mathrm{uniform}(-\beta_{\mathrm{dur}}, \beta_{\mathrm{dur}})\bigr),
\end{aligned}
\label{eq:ram_dur}
\end{equation}

where $m$ and $c$ are linear coefficient parameters controlling task size, $i$ is the chromosome identifier, and $\beta_{\text{ram}}$ and $\beta_{\text{dur}}$ are parameters controlling the magnitude of random noise. This noise, sampled from a uniform distribution, accounts for the natural variability in real genomic data.

In this simulation, an overcommitted task—one that exceeds its allocated RAM—is handled under a worst-case scenario assumption: it fails at the end of execution and must be rerun. This is equivalent to doubling its execution time. Any priors for the predictor are obtained from a single, noisy run of the same pipeline in order to test the scheduler's robustness in realistic conditions. 
The final parameters selected for the dynamic scheduler are: $s = 1.30$, $\gamma_{max} = 0.95$, $\gamma_{min}=0.80$, $p = 2$, and polynomial regression degree $d=1$. 




\subsubsection{Scheduler Module Evaluation.}

\begin{figure*}[htbp]
    \centering
    \includegraphics[width=1.0\textwidth]{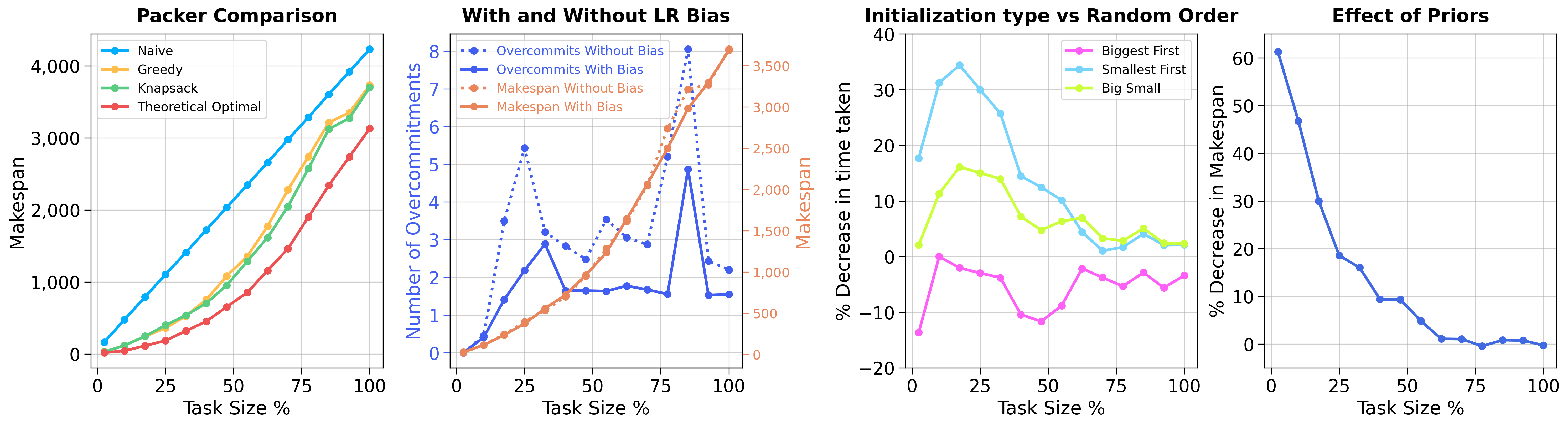} 
    \caption{\textbf{Scheduler Module Evaluation}: \textit{(Packer Comparison)} the knapsack packing produces the closest results to the theoretical limit. \textit{(With and Without LR Bias)} Including LR Bias showcases a decrease in overcommits, without affecting the makespan. \textit{(Initialization type vs Random Order)} the Smallest First initialization order produces the lowest makespan compared to the random initialization set. \textit{(Effect of Priors)} Effect of incorporating priors given a task size.
    }
    \label{fig:scheduler}
\end{figure*}

To understand the contribution of each component to the overall scheduler performance, we conduct a modular evaluation across different configurations. The key components analyzed are the packing algorithm (Knapsack vs. Greedy), the predictor type (Polynomial Regression with and without percentile bias), and the predictor initialization order (Biggest, Smallest, or Biggest and Smallest tasks first). Figure \ref{fig:scheduler} visually summarizes these findings.

Figure \ref{fig:scheduler} (Packer Comparison) shows the makespan (total run time) as a function of the task size defined as the size of chromosome 1 (largest chromosome) relative to the available RAM, in percentage. We define two baselines: "Naive" scheduling, which runs each task sequentially forming an upper bound for makespan, and "Theoretical" Optimal scheduling, which assumes perfect knowledge of task requirements beforehand and solves the problem through constraint optimization, forming the lower bound for makespan. 
The results (Figure \ref{fig:scheduler}, Packer Comparison) show that while a simple Knapsack and Polynomial Regression combination is nearly $35\%$ less efficient than the theoretical optimal, the Knapsack packing strategy consistently produces a lower makespan compared to the Greedy algorithm, which remains close but less optimal. This highlights the importance of maximizing RAM utilization over simply the number of jobs at a given time. 

Moreover, the addition of the Polynomial Linear Regression (LR) bias term resulted in a significant decrease in overcommitments by $38\%$ on average (Fig. \ref{fig:scheduler} With and without LR Bias). Remarkably, we achieve this reduction while also decreasing makespan by an average of about $0.5\%$. This demonstrates that a carefully chosen bias effectively reduces scheduling failures by minimizing over-allocation penalties.

Figure \ref{fig:scheduler} (Initialization type vs Random Order) shows the impact of different initialization orders for the predictor in the absence of prior knowledge. We find that the \textit{Smallest First} initialization produced the largest gains in speed for most task sizes, leading to the lowest overall makespan. This is a crucial insight: while the \textit{Biggest and Smallest} order provides a more robust initial predictor and reduces overcommits by almost $20\%$, the performance gains from quickly completing the sequential initialization phase with smaller tasks outweigh the benefits of a more accurate predictor. It must be noted that as the task size increases, scheduling converges to a sequential setting, decreasing the effect of initialization strategies.


Figure \ref{fig:scheduler} (Effect of Priors) shows the relative decrease of makespan when including prior information as a function of the task size. The experiments based on simulations show that having priors, even if noisy, removes the need for sequential predictor initialization, leading to a decrease of makespan, and is especially effective for task sizes below $50\%$ of total RAM, where initial concurrency plays a vital role. This diminishes as task size increases and concurrency drops. Intuitively, as the task size increases (i.e., chromosome 1 size approaches the available RAM), the number of concurrent jobs approaches 1, moving from parallel to sequential processing.


Finally, Table \ref{tab:scheduler_performance} summarizes the aggregate effects of each module, bringing the final average makespan over all task sizes about $13\%$ closer to the theoretical limit, while decreasing overcommitment by almost $77\%$. It also highlights that the makespan from our method outperforms the recently published memory manager Sizey \cite{bader2024sizey} on most task sizes. 


\begin{table}[h!]
\small
\centering
\begin{tabular}{|c|l|c|c|}
\hline
\textbf{Size \%} & \textbf{Scheduler} & \textbf{Makespan} & \textbf{Overcommits} \\
\hline
\multirow{5}{*}{10} & Knapsack & 117.15 & 0.45 \\
 & + LR Bias & 116.96 & 0.41 \\
 & + Smallest Init & 103.75 & 1.12 \\
 & + Prior & 57.11 & \textbf{0.00} \\
 & Sizey & \textbf{51.47} & \textbf{0.00} \\
\cline{2-4}
 & Theoretical Limit & 42.80 & 0.00 \\
\hline
\multirow{5}{*}{40} & Knapsack & 703.06 & 2.83 \\
 & + LR Bias & 723.16 & 1.65 \\
 & + Smallest Init & 671.26 & 2.03 \\
 & + Prior & \textbf{627.63} & 0.95 \\
 & Sizey & 648.04 & \textbf{0.68} \\
\cline{2-4}
 & Theoretical Limit & 452.40 & 0.00 \\
\hline
\multirow{5}{*}{70} & Knapsack & 2047.82 & 2.87 \\
 & + LR Bias & 2063.59 & 1.67 \\
 & + Smallest Init & 1995.82 & 1.73 \\
 & + Prior & \textbf{1989.39} & 0.77 \\
 & Sizey & 2156.14 & \textbf{0.67} \\
\cline{2-4}
 & Theoretical Limit & 1462.20 & 0.00 \\
\hline
\multirow{5}{*}{100} & Knapsack & 3701.68 & 2.20 \\
 & + LR Bias & 3689.64 & 1.55 \\
 & + Smallest Init & 3603.64 & 1.23 \\
 & + Prior & \textbf{3589.73} & \textbf{0.59} \\
 & Sizey & 3607.96 & 0.80 \\
\cline{2-4}
 & Theoretical Limit & 3130.00 & 0.00 \\
\hline
\end{tabular}
\caption{Scheduler Performance Evaluation.}
\label{tab:scheduler_performance}
\end{table}


\begin{figure*}[htpb]
    \centering
    \includegraphics[width=1.0\textwidth]{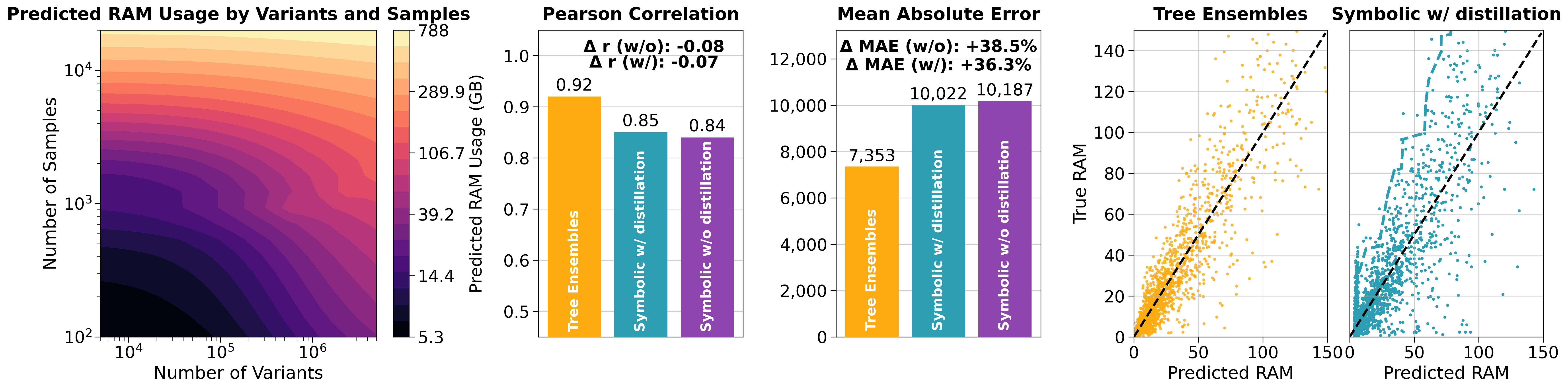}
    \caption{\textbf{RAM prediction results:}
    \emph{(Predicted RAM vs.\ variants and samples)} heatmap of $\widehat{y}$ for fixed $(V_{\mathrm{ref}},S_{\mathrm{ref}},\mathrm{Thr})$;
    \emph{(Pearson correlation)} teacher ensemble vs.\ symbolic regressor (with/without distillation);
    \emph{(Mean absolute error)} MAE at test set;
    \emph{(Tree Ensembles Scatter)} predicted vs.\ true RAM for the ensemble teacher;
    \emph{(Symbolic w/ distillation Scatter)} predicted vs.\ true RAM for the distilled symbolic model; dashed curve shows the 80th-percentile adjustment.}
    \label{fig:ram}
\end{figure*}

\begin{figure}[htpb]
    \centering
    \includegraphics[width=1.0\columnwidth]{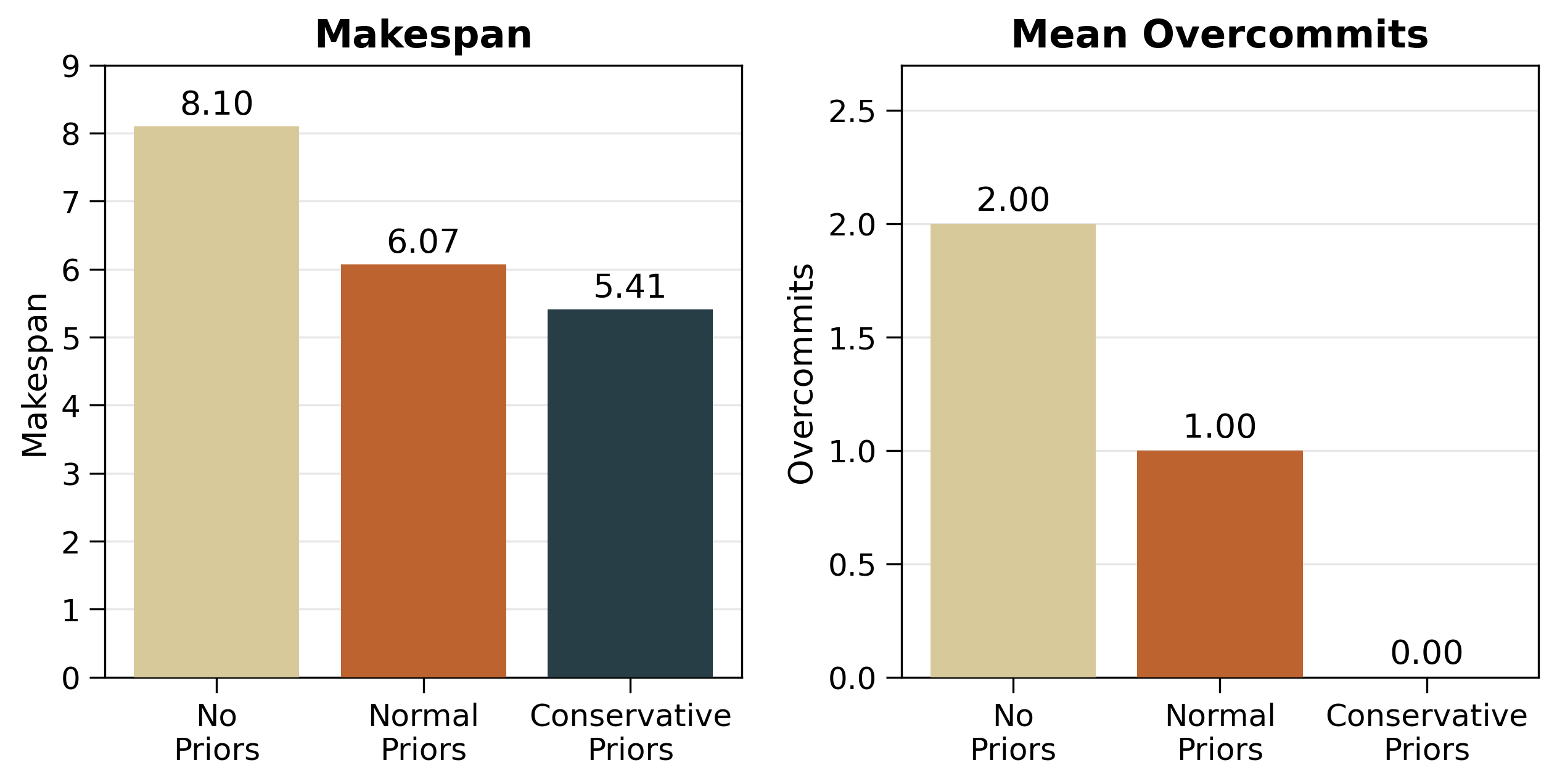} %
    \caption{\textbf{Deployed impact of conservative priors.} \textit{Beagle} makespan and overcommits in StrataRisk\textsuperscript{\texttrademark}. The calibrated priors nearly halves wall-clock time of the dynamic knapsack scheduler relative to the no-prior baseline.}

    \label{fig:ram-scheduler-deployed}
\end{figure}

\subsection{RAM Usage Prediction with Symbolic Regression}

Given the dataset composed of multiple runs of \textit{Beagle} with input files of different sizes and different software parameters, we train, calibrate, and evaluate the symbolic regression RAM prediction model.

\paragraph{Learned Expression.}
In the Beagle case study, the learned symbolic model takes the form:

\begin{equation}
\begin{aligned}
g(\tilde{\mathbf{x}}) &= c_0 \,\log\!\bigl( h_1(\tilde{\mathbf{x}})\,h_2(\tilde{\mathbf{x}}) \bigr) + c_5,\\[0.2em]
h_1(\tilde{\mathbf{x}}) &= \exp(\tilde{x}_5)\,
\log\!\Bigl(\sqrt{\Bigl\lvert
\frac{\tilde{x}_0 + c_1}{\,\tilde{x}_5 - \tfrac{\log\!\bigl(1+\exp(\tfrac{1}{2}\tilde{x}_4)\bigr)}{\,c_2 \tilde{x}_5 + c_3\,}}
\Bigr\rvert} + 1\Bigr),\\[0.2em]
h_2(\tilde{\mathbf{x}}) &= \bigl\lvert \tilde{x}_4 + \tilde{x}_6 + c_4 \bigr\rvert + 1
\end{aligned}
\label{eq:g_h1_h2}
\end{equation}

where $c_k$ are learned constants and $\tilde{x}_j$ denote standardized feature values. 
The learned expression reflects key drivers of RAM usage in genotype imputation. The number of samples $\tilde{x}_5$ in the exponential and divisor terms indicates an increase of memory required for storing haplotypes as sample counts grow. The combined effect of dataset size ($\tilde{x}_4$) and reference panel size ($\tilde{x}_6$) appears within additive and absolute value terms, capturing their joint contribution to memory requirements. The number of threads $\tilde{x}_0$ represents the effect of parallel execution on per-thread memory buffers.

Figure \ref{fig:ram} (Predicted RAM Usage by Variants and Samples) shows a heatmap of the predicted RAM values as the number of variants and samples changes (given a fixed set of reference panel and number of threads). Note that the RAM requirements grow from 5 to almost 800 GB as the number of variants ranges from around $10^4$ to $10^7$ and the number of samples ranges from $10^3$ to $10^4$.

\paragraph{Regression Evaluation.} Figure \ref{fig:ram} (Pearson Correlation, Mean Absolute Error) shows the correlation and error between predicted and ground truth RAM values from the Boosting Tree ensembles and the symbolic regression. Symbolic regression shows a minor decrease in performance compared to the Boosting Tree ensembles (Pearson correlation $0.92$ to $0.85$) while providing the benefits of an easy deployment. Furthermore, we can observe that using distillation instead of training the symbolic regression from scratch (without distillation) leads to performance improvement and an average error decrease.
Figure \ref{fig:ram} (Tree Ensembles, Symbolic w/ distillation) shows a scatter plot between predicted and true RAM, showing that on average the method accurately predicts the required RAM. The $80$th-percentile of the predictions (dashed line) provides a conservative estimate allowing a safer job scheduling without overcommitments.

\subsection{Improvements in Precision Medicine Workflows}

We integrated the proposed RAM–aware parallelization systems—static ordering, dynamic knapsack‐style packing, and the symbolic RAM predictor—into Galatea Bio's StrataRisk\textsuperscript{\texttrademark}, a clinically validated production pipeline that computes polygenic risk scores (PRS) directly from whole–genome sequencing. This integration is orthogonal to the scientific modeling stack (imputation, ancestry/LD corrections, cohort calibration), where the scheduler operates solely at the orchestration layer, packing chromosome–level tasks to maximize utilization while avoiding OOM failures.

Galatea Bio's StrataRisk\textsuperscript{\texttrademark} delivers PRS for a broad set of indications, spanning cardiometabolic (e.g., coronary artery disease, type 2 diabetes, lipid traits) and cancer phenotypes (e.g., breast and prostate cancer), with high predictive performance across diverse populations. By replacing the prior static queue with our dynamic scheduler, we observe a substantial reduction in end–to–end wall–clock makespan. Specifically, we monitor the makespan of the execution of \textit{Beagle} within the pipeline as it incorporates both the dynamic scheduler and the RAM prediction with symbolic regression. Figure \ref{fig:ram-scheduler-deployed} shows the makespan and overcommits of \textit{Beagle} when processing a patient's genome averaged across five runs. We observe a decrease of almost 2-fold in makespan when including the conservative RAM prediction with symbolic regression (conservative priors) within the dynamic scheduler. Note that the conservative prediction leads to zero overcommits. The improved scheduler translates into more efficient use of resources, lower compute cost per sample, and faster turnaround from sequence data to patient‐readable reports, enabling StrataRisk\textsuperscript{\texttrademark} to scale economically while delivering results to patients sooner.

\section{Conclusion}

We presented a framework for chromosome-level genomics that combines: (i) a static ordering to minimize peak memory under fixed concurrency, (ii) a dynamic knapsack scheduler with online RAM updates, and (iii) a compact symbolic RAM predictor distilled from ensembles. In simulations and ablations, the dynamic scheduler based on knapsack provided significant speedups, reducing the makespan and the number of overcommits while increasing concurrency. With symbolic regression priors, the sequential warm-up of the dynamic scheduler disappears and wall-clock drops at moderate task-to-RAM ratios. Deployed in a clinical PRS pipeline, the proposed approaches balance memory across threads, prevent OOM requeues, boost utilization, and lower cost. Prediction-guided, memory-aware scheduling is therefore a practical system for scaling precision-medicine genomics. Future directions include exploring more fine-grained parallelization such as simultaneously processing subsequences within a chromosome.

\section{Competing Interests}
CDB, FMDLV, AGI, and DMM have equity in Galatea Bio, Inc.; this work was conducted at Galatea Bio, Inc. and not as part of the authors' university duties or responsibilities.

\bibliography{references} 

\end{document}